\documentclass[12pt,a4paper]{article}
\usepackage{amssymb}

\begin{document}

\newcommand{\be}{\begin{equation}}
\newcommand{\ee}{\end{equation}}
\newcommand{\beann}{\begin{eqnarray*}}
\newcommand{\eeann}{\end{eqnarray*}}
\newcommand{\bea}{\begin{eqnarray}}
\newcommand{\eea}{\end{eqnarray}}
\newcommand{\lb}{\label}
\begin{titlepage}

\noindent

\vspace*{1cm}
\begin{center}
{\large\bf On the interpretation of quantum theory -- from Copenhagen to
            the present day} 

\vskip 1cm

{\bf Claus Kiefer} 
\vskip 0.4cm
Institut f\"ur Theoretische Physik,\\ Universit\"{a}t zu K\"oln, \\
Z\"ulpicher Str.~77,
50937 K\"oln, Germany.\\
\vspace{1cm}

\begin{abstract}
A central feature in the Copenhagen interpretation is the use of classical
concepts from the outset. Modern developments show, however, that the
emergence of classical properties can be understood within the
framework of quantum theory itself, through the process of decoherence.
This fact becomes most crucial for the interpretability of quantum
cosmology -- the application of quantum theory to the Universe as a whole.
I briefly review these developments and emphasize the importance of
an unbiased attitude on the interpretational side for future
progress in physics.
\end{abstract}
\end{center}

\end{titlepage}

\begin{quote}
Ich bin nicht damit zufrieden, wenn man eine Maschinerie hat,
die zwar zu prophezeien gestattet, der wir aber keinen klaren Sinn
zu geben verm\"ogen.
\vskip 3mm
Albert Einstein in a letter to Max Born (3.12.1953)
\end{quote}

\section{Copenhagen interpretations and alternatives}

A physical theory contains both a mathematical formalism and
an interpretational scheme. Although their relation may be
subtle already for classical physics, it becomes highly non-trivial
in quantum theory. In fact, although the mathematical framework
had been basically fixed by 1932, the debate about its meaning is going on.
As we shall see, this is due to the possibility of having conflicting
concepts of reality, without contradicting the formalism. This is why
all physicists agree on the practical application 
of the formalism to concrete problems such as the calculation
of transition probabilities.

It is often asserted that the orthodox view is given by the
Copenhagen interpretation of quantum theory. This is the standpoint
taken by most textbooks. What is this interpretation?
It is generally assumed that it originated from intense discussions
between Bohr, Heisenberg, and others in Copenhagen in the years 1925-27.
However, there has never been complete agreement about the
actual meaning, or even definition, of this interpretation even among
its main contributors. In fact, the Copenhagen interpretation has
remained until today an amalgamation of different views.
As has been convincingly argued in \cite{MB}, it is the incompatibility
between Bohr's and Heisenberg's views that sometimes gives the impression of
inconsistencies in the Copenhagen interpretation.

Historically, Heisenberg wanted to base quantum theory solely on
observable quantities such as the intensity of spectral lines,
getting rid of all intuitive ({\em anschauliche}) concepts such
as particle trajectories in space-time \cite {WH1}. This attitude
changed drastically with his paper \cite{WH2} in which he introduced
the uncertainty relations -- there he put forward the point of view that
it is the theory which decides what can be observed. His move from
positivism to operationalism can be clearly understood as a reaction
on the advent of Schr\"odinger's wave mechanics \cite{MB} which,
in particular due to its intuitiveness, became soon very popular among
physicists. In fact, the word {\em anschaulich} (intuitive) is contained
in the title of Heisenberg's paper \cite{WH2}.

Bohr, on the other hand, gave the first summary of his
interpretation in his famous lecture delivered in Como in September~1927,
cf. \cite{Como}. There he introduced the notion of complementarity
between particles and waves -- according to von Weizs\"acker the
core of the Copenhagen interpretation \cite{vW}.
As is well known, he later extended complementarity to non-physical
themes and advanced it to a central concept of his own philosophy. 
Complementarity means that quantum objects are neither particles
nor waves; for our intuition we have to use both pictures,
in which the range of applicability of one picture
necessarily constrains the range of applicability of the other.
Heisenberg, as a mathematical physicist, was not at ease with such an
interpretation. He preferred to use one coherent set of concepts,
rather than two imcompatible ones, cf. \cite{MB}. In fact, it was known
by then that particle and wave language can be converted into each other
and are transcended into the consistent formalism of quantum theory.
As Heisenberg wrote in his book \cite{WH3}: ``Licht und Materie sind
einheitliche physikalische Ph\"anomene, ihre scheinbare Doppelnatur
liegt an der wesentlichen Unzul\"anglichkeit unserer Sprache.
\ldots Will man trotzdem von der Mathematik zur anschaulichen
Beschreibung der Vorg\"ange \"ubergehen, so mu\ss\ man sich mit
unvollst\"andigen Analogien begn\"ugen, wie sie uns
Wellen- und Partikelbild bieten\footnote{``Light and matter
are unique physical phenomena, their apparent double nature
is due to the essential inadequacy of our language. \ldots
If one nevertheless wants to procede from the mathematics to
the intuitive description of the phenomena, we have to restrict
ourselves to incomplete analogies as they are offered by the wave and
particle pictures.''}.''

Even Bohr's own interpretation did change in the course of time. 
This happened in particular due to the influence of the
important paper by Einstein, Podolsky and Rosen in 1935 (EPR)
\cite{EPR}. Before EPR, an essential ingredient 
of his interpretation was the uncontrollable
disturbance of the quantum system by the apparatus during a measurement.
The analysis of EPR demonstrated, however, that the issue is not the
disturbance, but the non-separability (the entanglement) of a quantum system
over in principle unlimited spatial distances. Therefore, in his response
to EPR \cite{NB2}, Bohr adopted a strong operationalistic attitude,
concealing the crucial concept of entanglement.
Taking the indispensability of classical concepts for granted, he argued that
even without any mechanical disturbance there is an ``influence of the
very conditions which define the possible types of predictions regarding the 
future behavior of the system'' \cite{NB2},
i.e. no simultaneous reality for measurements of noncommuting variables
such as position and momentum should exist.
 This attitude leads eventually
to the consequence that quantum systems would not possess a real existence
before interacting with a suitable measurement device
(``only an observed phenomenon is a phenomenon'').

A somewhat ambigous role in the Copenhagen interpretation(s) is
played by the ``collapse'' or ``reduction'' of the wave function.
This was introduced by Heisenberg in his uncertainty paper
\cite{WH2} and later postulated by von Neumann as a dynamical
process independent of the Schr\"odinger equation, see Sec.~2. 
Most proponents of the Copenhagen interpretation have considered
this reduction as a mere increase of knowledge (a transition from
the potential to the actual), therefore denying that the wave function
is a kinematical concept and thus affected by dynamics.
 The assumption of a dynamical collapse
would definitely be in conflict with Bohr's ideas of complementarity
which forbid a physical analysis of the measurement process.
To summarise, one can identify the following ingredients  
as being characteristic for the Copenhagen interpretation(s):
\begin{itemize}
\item Indispensability of classical concepts for the description
      of the measurement process
\item Complementarity between particles and waves
\item Reduction of the wave packet as a formal rule without
      dynamical significance
\end{itemize}
This set of rules has been sufficient to apply the
quantum formalism pragmatically to concrete problems. 
But is it still sufficient? And is it satisfactory? 

Modern developments heavily rely on the concept of entanglement,
in order to describe satisfactorily the precision experiments that
are now being performed in quantum optics and other fields. 
 From the experimental violation of Bell's inequalities
 it has become evident, that quantum theory cannot be substituted
by a theory referring to a local reality. This is of course a
consequence of the superposition principle -- the heart of
quantum theory. 
 Among recent developments employing entanglement
are quantum computation and quantum cryptography
\cite{RMP} and the reversible transition from the (coherent) superfluid phase
to an (incoherent) Mott insulator phase in a Bose-Einstein condensate
\cite{BE}, during which interference patterns appear and disappear.
The superposition principle is indispensable for describing $K-\bar{K}$
and neutrino oscillations.
It would be hard to imagine how all this can be understood by denying any
dynamical nature of the wave function and to interpret it as describing
mere knowledge. Moreover, it is now clear,
both theoretically and experimentally, that the classical appearance
of our world can be understood as a dynamical process
{\em within} quantum theory itself, without any need to postulate it. 
Therein, entanglement plays the crucial role. This will be discussed
in the next section. 

There have been, in the course of history, various attempts to
come up with an alternative to the Copenhagen interpretation
\cite{MJ,WZ}. Here I want to mention two of them,
the de~Broglie-Bohm interpretation and the Everett interpretation.
Many other interpretations are some variant or mixture of these two
and the Copenhagen interpretation. The consistent-histories interpretation,
for example, contains elements from both Copenhagen and Everett,
see e.g. Chap.~5 in \cite{deco}. Different from these interpretations
are attempts which aim at an explicit change of the Schr\"odinger
equation in order to get a dynamical collapse of the wave function,
see e.g. Chap.~8 in \cite{deco}. Up to now, however, there is no experimental
hint that the Schr\"odinger equation has to be modified.

In the de~Broglie-Bohm interpretation
(or ``theory''), the wave function $\Psi$ is
supplemented with classical variables (particles and fields)
 possessing definitive values of
position and momentum. Whereas the wave function obeys an
autonomous dynamics (obeying the Schr\"odinger equation without
additional collapse), the particle dynamics depends on $\Psi$
(often called a guiding field).
Assuming that the particles are distributed according to
$\vert\Psi\vert^2$, the predictions of this theory are indistinguishable
from the ordinary framework, at least within non-relativistic
quantum mechanics. After a measurement, the particle is trapped
within one particular wave packet with the usual quantum-mechanical
probability. Because the other wave packets 
 are spatially separated from it after the measurement, they can no longer
influence the particle. This represents an apparent collapse of the
wave function and the occurrence of a definite measurement result.
 In principle, however, the remaining packets,
although empty, can interfere again with the packet containing the
particle, but the probability for this is tiny in macroscopic
situations. 

John Bell called the Everett interpretation 
a Bohm interpretation without trajectories.
In fact, Everett assumes just as Bohm that the wave function 
is part of reality
and that there is never any collapse.  
Therefore, after a measurement, all components corresponding
to the different outcomes are equally present. It is claimed that
the probability interpretation of quantum theory can be derived
from the formalism (which is, however, a contentious issue).
 Von Weizs\"acker
calls this interpretation \cite{vW} ``\ldots die einzige, die nicht hinter
das schon von der Quantentheorie erreichte Verst\"andnis
zur\"uck-, sondern vorw\"arts \"uber es 
hinausstrebt\footnote{``\ldots the only one that does not
fall back behind the understanding already achieved by quantum theory
but which strives forwards and even beyond.''}.''
The open question is of course when and how these different
components (``branches'') become independent of each other. 
This leads me to the central topic -- the emergence of classical
behaviour in quantum theory.

\section{The emergence of classical properties in quantum theory}

If classical concepts are not imposed from the outset,
they have to be derived from the formalism, at least in an
approximate sense. John von Neumann was the first who analysed in 1932 the
measurement process within quantum mechanics. He considers the coupling
of a system (S) to an apparatus (A), see Fig.~1. 

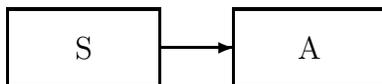
\begin{figure}[h]
\begin{center}
\setlength{\unitlength}{1cm}
\begin{picture}(5,1) \thicklines %\linethickness{1mm}
\put(0,0){\framebox(2,1){S}}
\put(3,0){\framebox(2,1){A}}
\put(2,0.5){\vector(1,0){1}}
\end{picture}
\end{center}
\caption{Original form of the von Neumann measurement model.}
\end{figure}

Let the states of the measured system which are
discriminated by the apparatus be denoted by $|n\rangle$
(for example, spin up and spin down), then an
appropriate interaction Hamiltonian has the form
(see Chap.~3 in \cite{deco}, or \cite{KJ})
\be H_{int} =\sum_n|n\rangle\langle n| \otimes\hat{A}_n\ .
\label{Hamil} \ee
The operators $\hat{A}_n$, acting on the states of the apparatus, are
rather arbitrary, but must of course depend on the ``quantum number'' $n$.
Eq. (\ref{Hamil}) describes an ``ideal'' interaction during which
the apparatus becomes correlated with the system state, without changing
the latter. There is thus no disturbance of the system by the
apparatus -- on the contrary, the apparatus is disturbed by the system
(in order to yield a measurement result).

If the
measured system is initially in the state $|n\rangle$ and the device in
some initial state $|\Phi_0\rangle$,
the evolution according to the Schr\"odinger equation
with Hamiltonian (\ref{Hamil}) reads
\bea |n\rangle|\Phi_0\rangle \stackrel{t}{\longrightarrow}
     \exp\left(-{\rm i} H_{int}t\right)|n\rangle|\Phi_0\rangle
     &=& |n\rangle\exp\left(-{\rm i}\hat{A}_nt\right)|\Phi_0\rangle\nonumber
\\
     &=:& |n\rangle|\Phi_n(t)\rangle\ .  \label{ideal} \eea
The resulting apparatus states $|\Phi_n(t)\rangle$ are often called
``pointer states''.
A process analogous to (\ref{ideal}) can also be
formulated in classical physics. The essential new quantum features
now come into play when we consider a {\em superposition} of different
eigenstates (of the measured ``observable'') as initial state. The
linearity of time evolution immediately leads to
\be \left(\sum_n c_n|n\rangle\right)|\Phi_0\rangle
    \stackrel{t}\longrightarrow\sum_n c_n|n\rangle
    |\Phi_n(t)\rangle\ . \label{measurement}
\ee
But this state is a superposition of macroscopic measurement results
(of which Schr\"odinger's cat is just one drastic example)!
To avoid such a bizarre state, and to avoid the apparent conflict with
experience, von Neumann introduced the dynamical collapse of the
wave function as a new law. The collapse should then select one component
with the probability $\vert c_n\vert^2$. He even envisaged
that the collapse is eventually caused by the consciousness of 
a human observer, an interpretation that was later also adopted by Wigner.
In the Everett interpretation, all the branches (each component
in (\ref{measurement})) are assumed to co-exist simultaneously.

Can von Neumann's conclusion and the introduction of the collapse
be avoided?
The crucial observation \cite{Zeh70}, which enforces an
extension of von Neumann's measurement theory, is the fact
that macroscopic objects (such as measurement devices) are so strongly
coupled to their natural environment, that a unitary treatment
as in (\ref{ideal}) is by no means sufficient and has to be modified
to include the environment.

\begin{figure}[ht]
\begin{center}
\setlength{\unitlength}{1cm}
\begin{picture}(10,2)(-0.5,-0.5) \thicklines %\linethickness{1mm}
\put(0,0){\framebox(2,1){S}}
\put(3,0){\framebox(2,1){A}}
\put(2,0.5){\vector(1,0){1}}
\put(6,-0.5){\framebox(3,2){E}}
\put(5,0.3){\vector(1,0){1}}
\put(5,0.5){\vector(1,0){1}}
\put(5,0.7){\vector(1,0){1}}
\put(-0.5,-0.5){\dashbox{0.2}(6,2){}}
\end{picture}
\end{center}
\caption
{Realistic extension of the von Neumann
 measurement model including the environment. Classical properties 
emerge through the unavoidable, irreversible interaction
 of the apparatus with the environment.}
\end{figure}
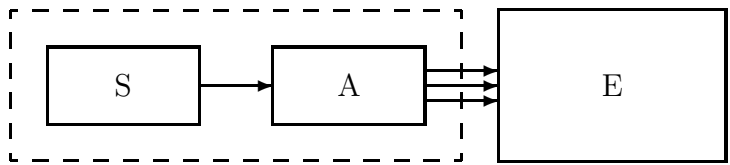

Fortunately, this can easily be done
to a good approximation, since the interaction with the environment
has in many situations the same form  as given by the Hamiltonian
(\ref{Hamil}): The measurement device is itself ``measured''
(passively recognised) by the environment, according to
\be \left( \sum_n c_n|n\rangle|\Phi_n\rangle\right)|E_0 \rangle
  \quad  \stackrel{t}\longrightarrow \quad
   \sum_n c_n|n\rangle |\Phi_n\rangle |E_n\rangle .\label{ideal2}
\ee
This is again a macroscopic superposition, now including the
myriads of degrees of freedom pertaining to the environment
(gas molecules, photons, etc.). However, most of these
environmental degrees of freedom are inaccessible.
Therefore, they have to be integrated out from the full state
(\ref{ideal2}). This leads to the reduced density matrix
for system plus apparatus, which contains all the information
that is available there. It reads
\be
\rho_{SA} \approx \sum_n |c_n|^2 |n\rangle\langle n|
   \otimes |\Phi_n \rangle \langle \Phi_n |
   \qquad\mbox{if}\qquad
   \langle E_n|E_m \rangle \approx \delta_{nm}\ ,
 \label{deco}
\ee
since under realistic conditions, different environmental states
are orthogonal to each other. 
Eq.~(\ref{deco}) is identical to the
density matrix of an ensemble of measurement results
$|n\rangle |\Phi_n\rangle$. System and apparatus thus seem to
be in one of the states $|n\rangle$ and $|\Phi_n\rangle$, given by
the probability $\vert c_n\vert^2$. 

Both system and apparatus thus assume classical properties
through the unavoidable, irreversible interaction with the
environment. This dynamical process, which is fully described
by quantum theory, is called {\em decoherence} \cite{deco}.
It is based on the quantum entanglement between apparatus
and environment.
Under ordinary macroscopic situations, decoherence occurs on an extremely
short timescale, giving the impression of an instantaneous collapse
or a ``quantum jump''. Recent experiments were able to demonstrate 
the continuous emergence of classical properties in mesoscopic systems
\cite{SH,DW}. Therefore, one would never ever be able to observe
a weird superposition such as Schr\"odinger's cat, because 
the information about this superposition would almost instantaneously
be delocalised into unobservable correlations with the
environment, resulting in an apparent collapse for the cat state. 
The concept of decoherence motivated Wigner to give up his
explanation of the collapse as being caused by consciousness \cite{EW}.
In fact, decoherence makes it evident that living creatures play no
particular role in the interpretation of quantum theory. 

The interaction with the environment distinguishes the local basis
with respect to which classical properties (unobservability of
interferences) hold. This ``pointer basis'' must obey the condition of
robustness, i.e. it must keep its classical appearance over the
relevant timescales \cite{Zurek,Zeh1}. Classical properties are thus not
intrinsic to any object, but only defined by their interaction with
other degrees of freedom. In simple (Markovian) situations
the pointer states are given by localised Gaussian states \cite{DK}.
They are, in particular, relevant for the localisation of
macroscopic objects.

To summarise, these developments have shown that classical concepts
are not an indispensable input to the theory,
as required by the Copenhagen interpretation, but a natural
consequence of the theory itself when applied to realistic conditions.

\section{Quantum gravity}

The modern developments discussed in the last section show that
the main assumptions of the Copenhagen interpretation, such as
complementarity and the demand for a priori classical concepts,
are not obligatory. Still, one might argue, it could be
possible to adopt this interpretation as a convenient background
for pragmatic use. 

While this may be true for ordinary laboratory situations, it may
become impossible if quantum effects of the gravitational field
are involved. It must be admitted that no effects of quantum gravity
have yet been seen, or identified as such, and that no final consensus
about such a theory has emerged. However, there exist many models
with important applications in cosmology; without an appropriate
and consistent interpretation, such models would be void of interest.

In ordinary quantum theory, time is given as an external parameter
and not subject to quantisation. On the other hand, the gravitational
field is described by Einstein's theory of general relativity,
in which space and time are dynamical and not absolute. I have discussed
elsewhere the reasons why one generally believes that gravity must
be described by a quantum theory at the most fundamental level
\cite{Karpacz}. If one quantises a theory that classically possesses
no absolute time -- be it general relativity or some alternative
theory -- the ensuing quantum theory does not contain any time
parameter at all. Since the role of time in general relativity is merely
to parametrise spacetimes (the ``trajectories'' in the corresponding
configuration space), the absence of time in quantum gravity is the
consequence of the absence of trajectories in quantum theory.
The central equation of quantum gravity is of the general form
\be
\hat{H}\Psi=0\ , \label{wdw}
\ee
where $\hat{H}$ is the total Hamilton operator of both gravitational
and non-gravitational fields. The total quantum state is thus of a
stationary form. Since quantum degrees of freedom are very sensitive
to their environment (see Sec.~2), and since the dominant interaction
in the Universe on large scales is gravity, one is immediately led 
to consider a quantum theory of the whole Universe --
quantum cosmology \cite{Karpacz}.
Since the Universe naturally contains all of its
observers, the problem arises to come up with an interpretation
of quantum theory that contains no classical realms on the
fundamental level.

How can a temporal dynamics be understood from the stationary equation
(\ref{wdw})? It has been demonstrated that a concept of time
re-emerges in a semiclassical limit 
as an approximate concept \cite{Karpacz,Zeh1}. In this limit,
an effective time-dependent Schr\"odinger equation holds along formal
WKB trajectories. Since different semiclassical branches usually decohere
from one another, an observer cannot experience the other branches
which only together form the one wave function $\Psi$ in (\ref{wdw}).

Clearly, the Copenhagen interpretation cannot cope with
quantum cosmology. This is the reason why most people working in this
field, at least implicitly, adopt the Everett interpretation, since
it is hard to imagine from where a conceivable collapse could emerge.
The problems that are addressed in quantum cosmology include
the quantum origin of the Universe, the quantum probability for the
occurrence of an inflationary phase, and the quantum-to-classical
transition for primordial fluctuations which serve as the seeds for
structure formation in the Universe, giving rise to galaxies
\cite{Karpacz}. With the advent of precision measurements for
the spectrum of the cosmic microwave background radiation, some of these
questions gain observational significance. It would thus be
unsatisfactory to avoid a theoretical description of such processes
just because some pre-conceived interpretation of quantum theory does 
not fit this purpose.

\section{Conclusion}

The Copenhagen interpetation needs classical concepts as
prerequisites. On the other hand, quantum theory itself predicts the
occurrence of decoherence through which systems
such as measurement devices can appear classically
to local observers. This can be, and has been, interpreted as
a quantum justification for at least part of the original Copenhagen
programme \cite{Zurek}. This explains, in retrospect, why the
Copenhagen interpretations can serve as a background for pragmatic
calculations, at least in non-gravitational situations.

The process of decoherence is based on the validity of the Schr\"odinger
equation and can thus not describe any real collapse. For local
experiments, this is definitely not needed, because decoherence
can explain why we seem to observe a collapse or a ``quantum jump''.
It would seem unnatural and ad hoc to introduce a real collapse
at this stage, where an apparent collapse is predicted
anyway by the Schr\"odinger equation. 
The concept of quantum jumps thus plays the role of epicycles in
astronomy \cite{ES1} -- it describes naively what is observed,
but it becomes redundant at the fundamental, theoretical, level.

Still, the measurement problem is not resolved 
for the total system including the
environment. The only alternatives so far are to either assume an
additional real collapse in violation of the Schr\"odinger equation
(for which there is not yet any experimental hint), or to adopt
an Everett interpretation with its simultaneous reality
of all branches. It seems hard to imagine that an experimental
decision between these alternatives can be made in the foreseeable
future. However, it is important to keep an open mind and to avoid
the burden of a pre-imposed interpretation -- ``sonst w\"are ernstlich
zu bef\"urchten, da\ss\ es dort, wo wir das Weiterfragen verbieten, wohl doch
noch einiges Wissenswerte zu fragen gibt'' \cite{ES2}.\footnote{``Otherwise
it would be seriously feared that just there, where we forbid further 
questions, there might still be something worth knowing that we
might ask about.''}

\end{document}